\documentclass[galaxies,article,accept,pdftex,moreauthors]{Definitions/mdpi} 

\usepackage[final]{changes}

\definecolor{color1}{rgb}{0, 0.5, 1.0}
\definecolor{grey}{rgb}{0.7, 0.7, 0.7}
\definecolor{ro}{rgb}{1, 0.7, 0.7}

\newcommand{\jkas}{J. Korean Astron. Soc.}
\newcommand{\apj}{Astrophys. J.}
\newcommand{\apjs}{Astrophys. J., Suppl. Ser.}
\newcommand{\apjl}{Astrophys. J. Lett.}
\newcommand{\aj}{Astron. J.}
\newcommand{\mnras}{Mon. Not.
R. Astron. Soc.}
\newcommand{\aap}{Astron. Astrophys.}

\newcommand{\pasj}{Publ. Astron. Soc. Jpn.}

\newcommand{\nar}{New Astron. Rev.}

\newcommand{\apss}{Astrophys. Space Sci.}
\newcommand{\raa}{Res. Astron. Astrophys.}

\firstpage{1} 
\pubvolume{1}
\issuenum{1}
\articlenumber{0}
\pubyear{2023}
\copyrightyear{2023}
\externaleditor{Academic Editor: Luigina Feretti} 
\datereceived{25 December 2022} 
\dateaccepted{06 February 2023} 
\datepublished{} 
\hreflink{https://doi.org/} 

\Title{Transverse Oscillations of the M87 Jet Revealed by KaVA Observations}

\TitleCitation{Transverse Oscillations of the M87 Jet Revealed by KaVA Observations}

\Author{Hyunwook Ro 
 $^{1,2,}$*\orcidA{}, Kunwoo Yi $^{3}$, Yuzhu Cui $^{4,{5}}$, Motoki Kino $^{{6,7}}$, Kazuhiro Hada $^{{8, 9}}$, Tomohisa Kawashima $^{{10}}$, Yosuke~Mizuno~$^{4,{11,12}}$, Bong Won Sohn $^{1,2,{13}}$ and Fumie Tazaki $^{{14}}$\orcidB{}}

\AuthorNames{Hyunwook Ro, Kunwoo Yi, Yuzhu Cui, Motoki Kino, Kazuhiro Hada, Tomohisa Kawashima, Yosuke Mizuno, Bong Won Sohn and Fumie Tazaki}

\AuthorCitation{Ro, H.; Yi, K.; Cui, Y.; Kino, M.; Hada, K.; Kawashima, T.; Mizuno, Y.; Sohn, B.W.; Tazaki, F. } 

\address{%
$^{1}$ \quad Department of Astronomy, Yonsei University, Yonsei-ro 50, Seodaemun-gu, Seoul 03722, Republic of Korea; bwsohn@kasi.re.kr (B.{W}.S.)\\
$^{2}$ \quad Korea Astronomy \& Space Science Institute, Daedeokdae-ro 776, Yuseong-gu, Daejeon~34055,~Republic~of~Korea\\
$^{3}$ \quad Department of Physics and Astronomy, Seoul National University, Gwanak-gu, Seoul~08826,~Republic~of~Korea; kunwoo.yi@snu.ac.kr\\
$^{4}$ \quad Tsung-Dao Lee Institute, Shanghai Jiao Tong University, Shanghai 201210, China; {yuzhu\_cui@sjtu.edu.cn}~(Y.C.); mizuno@sjtu.edu.cn~(Y.M.)\\
$^{5}$ \quad {Zhejiang Lab, Hangzhou, Zhejiang 311121, China}\\
$^{{6}}$ \quad Kogakuin University of Technology \& Engineering, Academic Support Center, 2665-1 Nakano, Hachioji, Tokyo 192-0015, Japan; motoki.kino@nao.ac.jp\\
$^{{7}}$ \quad National Astronomical Observatory of Japan, 2-21-1 Osawa, Mitaka, Tokyo 181-8588, Japan\\
$^{{8}}$ \quad Mizusawa VLBI Observatory, National Astronomical Observatory of Japan, 2-12 Hoshigaoka, Mizusawa, Oshu, Iwate 023-0861, Japan; kazuhiro.hada@nao.ac.jp\\
$^{{9}}$ \quad Department of Astronomical Science, The Graduate University for Advanced Studies (SOKENDAI), 2-21-1~Osawa, Mitaka, Tokyo 181-8588, Japan\\
$^{{10}}$ \quad Institute for Cosmic Ray Research, The University of Tokyo, 5-1-5 Kashiwanoha, Kashiwa, Chiba~277-8582,~Japan; kawshm@icrr.u-tokyo.ac.jp\\
$^{{11}}$ \quad School of Physics and Astronomy, Shanghai Jiao Tong University, Shanghai 200240, China\\
$^{{12}}$ \quad Institut f\"{u}r Theoretische Physik, Goethe Universit\"{a}t, Max-von-Laue Str. 1, D-60438~Frankfurt~am~Main,~Germany\\
$^{{13}}$ \quad
Department of Astronomy and Space Science, University of Science and Technology, Yuseong-gu, Daejeon 34113, Republic of Korea 
 \\
$^{{14}}$ \quad Tokyo Electron Technology Solutions Limited, Iwate 023-1101, Japan; fumie.tazaki@gmail.com\\
}

\corres{Correspondence: hwro@yonsei.ac.kr}

\abstract{
Recent VLBI monitoring has found transverse motions of the M87 jet.
However, due to the limited cadence of previous observations, details of the transverse motion have not been fully revealed yet.
We have regularly monitored the M87 jet at KVN and VERA Array (KaVA) 22\,GHz from December 2013 to June 2016.
The average time interval of the observation is $\sim$~0.1 year, which is suitable for tracking short-term structural changes.
From these observations, the M87 jet is well represented by double ridge lines in the region 2
{--}
12 mas from the core.
We found that the ridge lines exhibit transverse oscillations in all observed regions with a\added{n} \added{average} period of \replaced{$0.94\pm0.12$}{$\sim1$}~years.
When the sinusoidal fit is performed, we found that the amplitude of this oscillation is \added{an order of} $\sim0.1$ mas, and the oscillations in the northern and southern limbs are almost in phase. Considering the amplitude, it does not originate from Earth's parallax.
We propose possible scenarios of the transverse oscillation, such as the propagation of jet instabilities \added{or magneto-hydrodynamic (MHD) waves} or perturbed mass injection around magnetically dominated accretion flows.}


\begin{document}




\section{Introduction}

M87 is a giant elliptical galaxy at a distance of 16.8 Mpc~\cite{ehtc19}, which harbors a supermassive black hole of $M_{\bullet}=6.5 \pm 0.7 \times10^9$ $M_{\odot}$~\cite{ehtc19} and a prominent jet that extends several kiloparsecs (kpc) away from the galaxy.
At this distance, an angular size of 1 milli-arcsecond (mas) corresponds to $\approx$ 0.081\,parsecs (pc) $\approx$ 130 Schwarzschild radius ($r_{\rm s}$), and a proper motion of 1\,mas/yr is equivalent to $\sim$~0.264\,c.
It is, therefore, one of the best targets for studying the detailed structure of (sub)pc-scale jets and their evolution over time.

The launching region of the M87 jet is being intensively studied by Very Long Baseline Interferometry (VLBI), from which there is increasing evidence that the jet accelerates at distances less than $\sim 10^6~r_{\rm s}$~\cite{asada14, mertens2016, hada17, park19}.
In addition, high-resolution VLBI observations investigated the edge-brightened structure of the M87 jet in the same region~\cite{asada2012, hada16, kim18, Nakamura2018, walker2018} and found that the M87 jet is collimating in a parabolic shape.
The co-existence of acceleration and collimation in the M87 jet is as predicted from the magneto-hydrodynamic (MHD) acceleration model, e.g., \cite[][]{Nakamura2018,mckinney06,komissarov2007,chatterjee19}.

\subsection*{Transverse Motions in the Parsec-Scale Structures of the M87 Jet} 
Several previous VLBI studies found that the pc-scale structure of the M87 jet exhibits transverse motions (i.e., the motion in the direction perpendicular to the jet axis). 
\citet{walker2018} have reported a quasi-periodic sideways shift in the M87 jet by analyzing roughly annual observations at a Very Long Baseline Array (VLBA), 43\,GHz, over 17 years from 1999 to 2016. 
The jet shows a quasi-periodic sideways shift of about 8--10 years, and the shift propagates outward with a speed significantly slower than the flow speed.
They concluded that the non-ballistic propagation speed of the long-term patterns is consistent with the propagation speed of the Kelvin--Helmholtz instability.
{\citet{britzen2017} have investigated the evolution of the ridge lines of the M87 jet using 31 VLBA observations at 15\,GHz, spanning a time range between July 1995 and May 2011.
They found that the M87 jet structures switch between two phases. 
In the first phase, the jet ridge lines are at least double, or the jet axis is displaced vertically. 
In the second phase, the jet ridge lines remain almost straight but smoothly curved, and the jet components are aligned along a jet axis. 
They suggested the transition period between the two phases is $\sim$~2 years. They proposed the origin as a turbulent mass loading into the jet from the accretion disk.}

Thus far, two different studies of the transverse motions of the M87 jet have been reported.
However, the periods of transverse motions found in both studies are quite different from each other, even though they conducted observations over a long period of more than 15~years.
This discrepancy means they are likely seeing different motions of the jets, implying that the transverse motions of the M87 jet may be more complex.
However, both studies are limited in fully describing the transverse motions of the jet due to the low cadence and non-uniform spacing of the observations.
Therefore, high-cadence monitoring with more regular time intervals is required to clearly understand the transverse motions in the M87~jet.

KaVA (KVN and VERA Array;~\cite{niinuma2014}) has regularly observed M87 since late 2013. 
The major goal was to detect the velocity profile of the M87 accurately. 
The pilot observations of KaVA 22\,GHz from December 2013 to June 2014 have successfully found superluminal motions in the region within 10\,mas from the core~\cite{hada17}. 
A more complete picture of the jet acceleration profile was revealed with the biweekly observations in 2016 at 22 and 43\,GHz~\cite{park19}.
KaVA monitoring of the M87 jet has the advantage of seeing the extended jet at shorter time intervals than the previous studies. 
In this work, we will use multi-epoch KaVA data to investigate the transverse motion of the M87 jet in detail.
In Section~\ref{sec2}, we describe the KaVA 22\,GHz monitoring data from December 2013 to June 2016. Parts of these data were published in previous studies, and some are new in this study.
In Section~\ref{sec3}, we present a method for creating ridge lines in the jet and displaying the distribution of ridge lines for the M87 jet of all data.
Section~\ref{sec4} analyzes the transverse motion of the M87 jet by fitting a sinusoidal curve and creating a periodogram. As a result, we found an oscillation with a period of $\sim$~1 year at all distances of the jet. We propose possible origins of transverse oscillation in the M87 jet.

\section{Observations}\label{sec2}

We used KaVA 22\,GHz observations from December 2013 to June 2016. Some of these data have already been published: 10 epochs from 5 December 2013 to 14 June 2014 in \citet[][]{hada17}, and 8 epochs of biweekly monitoring from 25 February 2016 to 13 June 2016 in \citet[][]{park19}. 
Data from 1 September 2014 to 16 May 2015 were newly added to this study after performing a standard data post-correlation process with the NRAO's Astronomical Image Processing System ({\tt AIPS};~\cite{greisen2003}). 
The detailed process is described by \citet{park19}. During this process, the amplitude of data was multiplied by a factor of 1.35 for the data before 2015 May and 1.3 after 2015 May to correct for amplitude loss during correlation~\cite{lee2015}. After that, imaging with CLEAN and self-calibration was performed using the {\tt Difmap} software package~\cite{shepherd1997}. When natural weighting is applied, the typical \replaced{Full Width at Half Maximum (FWHM)}{size of the FWHM} of the synthesized beam is close to a circular shape with radii of 1.2\,mas.
We present naturally weighted CLEAN images for the data from 1 September 2014 to 16 May 2015 in Figure~\ref{fig:total_intensity_map}.

In total, we used 24 KaVA 22\,GHz data to analyze the transverse motion. The image information for the data is summarized in Table~\ref{tab.a.1}.
Due to the relatively short on-source time, the dynamic ranges for observations in 2013$-$2015 are relatively low compared to observations in 2016.
However, it is possible to investigate jet structures up to $\sim$~12\,mas in almost all observations.
Compared to the observations used in \citet{walker2018} and \citet{britzen2017}, our KaVA 22\,GHz monitoring is not a long period of observation ($\sim$~2.52\,years).
However, due to its much shorter \replaced{cadence}{time interval} (on average $\sim$~0.1\,year), it is suitable for studying the transverse oscillations of jets in more detail.

\section{The Ridge Lines}\label{sec3}

The transverse motion of the M87 jet can be studied by tracking changes in the ridge lines.
Various definitions of ridge lines of AGN jet have been used in previous studies, e.g., \cite[][]{britzen2017, perucho2012, fromm13, cohen15, pushkarev17}.
To obtain the ridge lines of the M87 jet, the following steps were taken, as summarized in Figure~\ref{fig:Gaussian_model_fit_example}. 
Firstly, we restored all the maps to the same circular beam of 1.2\,mas~$\times$~1.2\,mas, which is the comparable size of the synthesized beam of KaVA at 22\,GHz. 
Then, we rotated all images by $-$18\,degrees in order to align the jet's central axis with the horizontal axis, e.g.,~\cite[][]{walker2018}.
We then obtained one-dimensional brightness distributions perpendicular to the jet axis using the {\tt AIPS} task \textit{SLICE} for all successive steps of 0.2\,mas down the jet starting at the core.
The M87 jet shows a complex jet structure with a prominent limb-brightening; hence, there is usually more than one peak in the transverse brightness distribution, e.g.,~\cite[][]{walker2018, kim18}.
However, due to limited angular resolution, the transverse slice plots of our observations show a single peak up to a certain distance from the core and double peaks for the downstream jet.
We fitted a single or double Gaussian to the transverse brightness distribution with the least squared method. 
Then, the ridge lines of the jet were derived by connecting the peaks of the Gaussian along the jet. 
We found that the boundary between single-ridge and double-ridge regions differs slightly from epoch to epoch depending on the data quality.
However, from 2\,mas from the core, the jet structure was sufficiently resolved into double ridges in all epochs.
Therefore, in further analysis, we set the boundary as 2\,mas for all epochs.

Figure~\ref{fig:Gaussian_model_fit_example} shows the ridge line of the M87 jet from one epoch of observation (25 February 2016).
The bottom panels show transverse slice plots at 1.8, 2.4, 6.0, and 9.2\,mas distance from the core. The solid black  line is the transverse brightness distribution, and the solid blue and solid green lines are the single and double Gaussian fitting results, respectively.
The error of the Gaussian peak \replaced{intensity}{flux} ($\sigma_{\text{peak}}$), peak position ($\sigma_{\text{position}}$) and size ($\sigma_{\text{width}}$) is calculated using Equations~\eqref{eq1},~\eqref{eq2}, and~\eqref{eq3}~\cite{fomalont99, lee08}: 

\begin{figure}[H]
    \captionsetup[subfigure]{position=top,textfont=normalfont,singlelinecheck=off,justification=raggedright}
    \begin{minipage}[b]{\linewidth}
        \centering
        \includegraphics[width=0.7\textwidth]{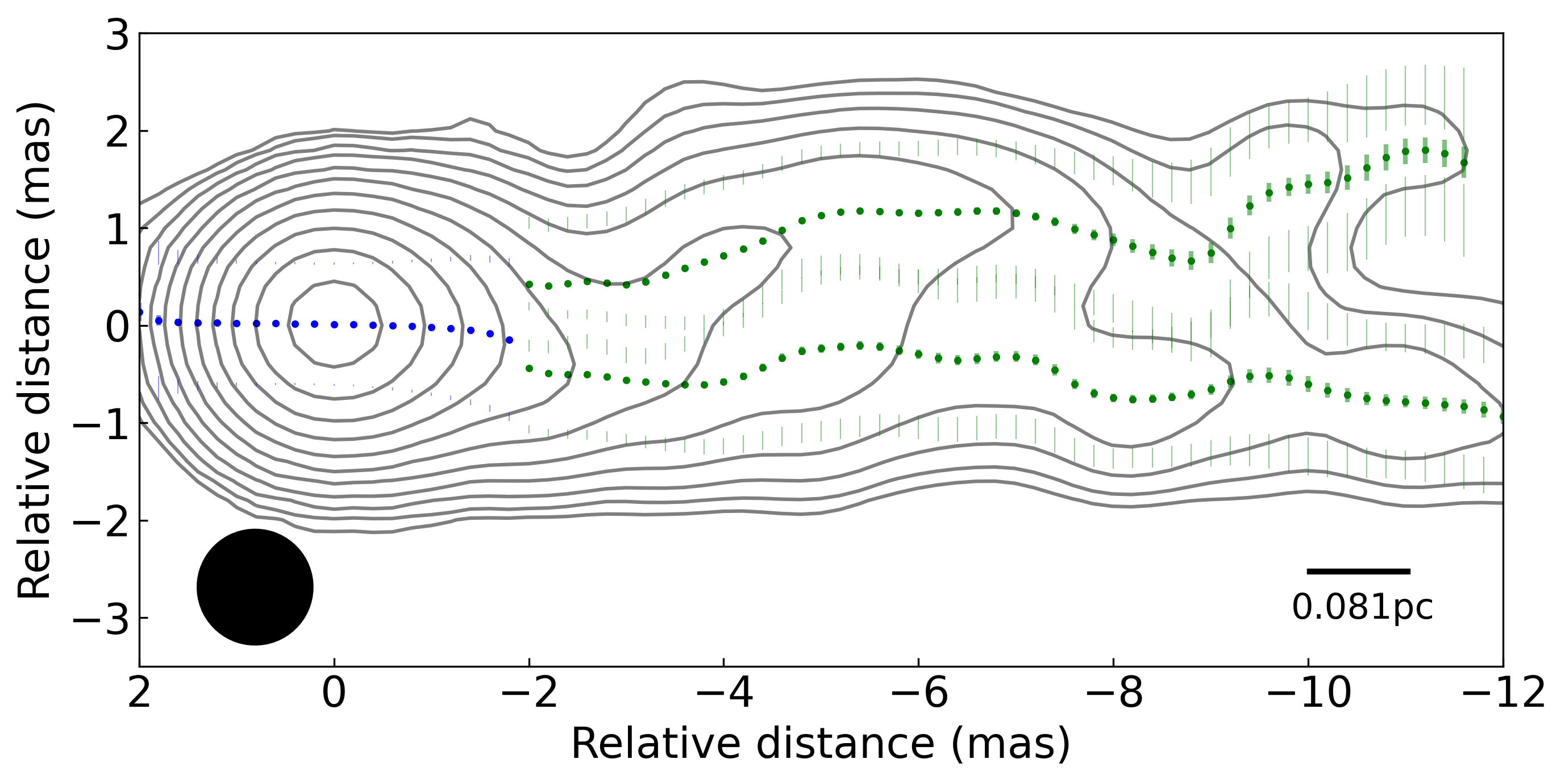}
    \end{minipage}
    \newline
    \begin{minipage}[b]{\linewidth}
        \centering
        \includegraphics[width=0.235\textwidth]{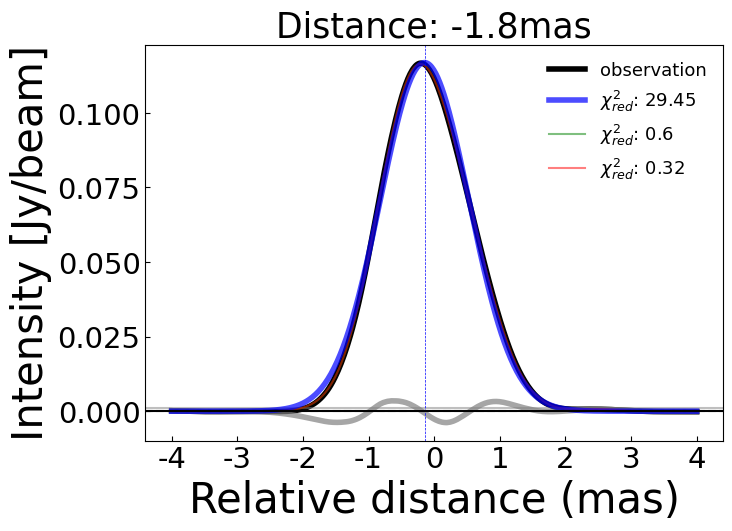}
        \includegraphics[width=0.235\textwidth]{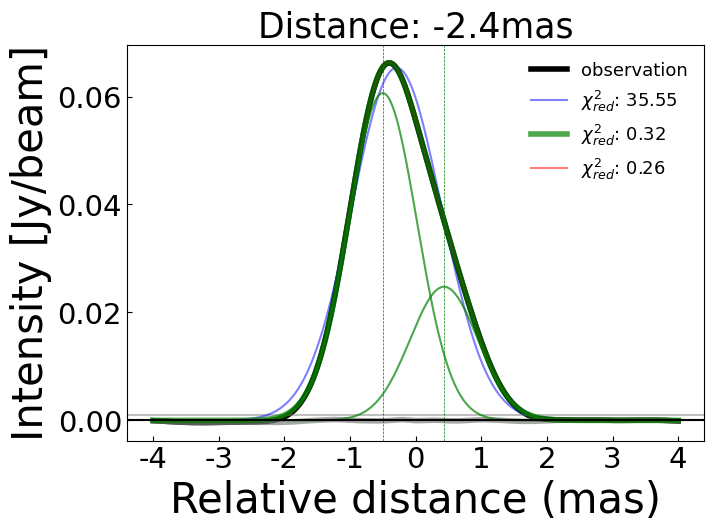}
        \includegraphics[width=0.235\textwidth]{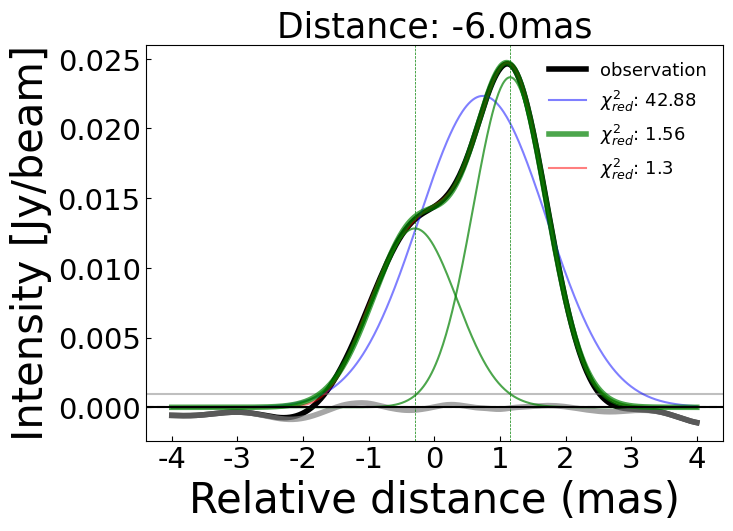}
        \includegraphics[width=0.235\textwidth]{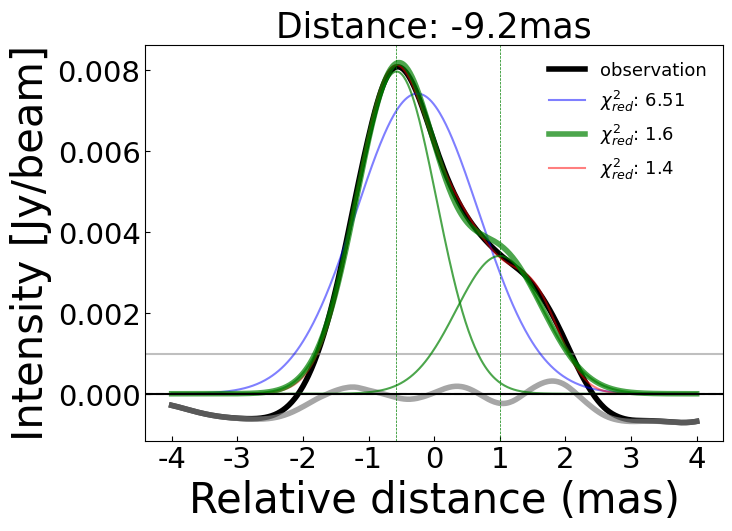}
    \end{minipage}
    \caption{\textbf{Top:} Example 
 of the Gaussian model fitting results for drawing the ridge line of one epoch at the KaVA 22\,GHz (25 February 2016). 
 The image has been rotated by {$-$}18 degrees. Contours start at 3$\sigma_{\text{rms}}$, increasing in steps of 2. The black circle in the bottom left corner represents the size of the restoring beam, namely 1.2\,mas circular Gaussian. \textbf{Bottom:} Transverse brightness distributions at 4 different locations (at distances 1.8, 2.4, 6.0, 9.2 mas) and their Gaussian model fit results. The blue and green dots are the peak and FWHM positions of the best-fit results of single or double Gaussian models,~respectively.}
    \label{fig:Gaussian_model_fit_example}
\end{figure}

\begin{equation}
    \sigma_{\text{peak}} = \sigma_{\text{rms}} \times \left(1+\frac{S_{\text{peak}}}{\sigma_{\text{rms}}}\right)^{1/2}\label{eq1}
\end{equation}
\begin{equation}
    \sigma_{\text{position}} = \sigma_{\text{peak}} \times \frac{d_{\text{FWHM}}}{2S_{\text{peak}}}\label{eq2}
\end{equation}
\begin{equation}
    \sigma_{\text{width}} = 2\sigma_{\text{position}}\label{eq3}
\end{equation}where $\sigma_{\text{rms}}$ is the image rms noise, $S_{\text{peak}}$ is the peak intensity of the Gaussian, and $d_{\text{FWHM}}$ is the width of the Gaussian.

In the top panel, the positions of the Gaussian peaks are distributed as blue dots in the inner 2\,mas from the core and as two green dots for more downstream. Errors in position are indicated by thick vertical lines. The Gaussian width is indicated by a thin vertical line, and the length of the line represents the error.
It can be seen that the position and width of the Gaussian model are smoothly connected, confirming the good tracking of the jet~structure.

\begin{figure}[H]
    \includegraphics[width=0.3\textwidth]{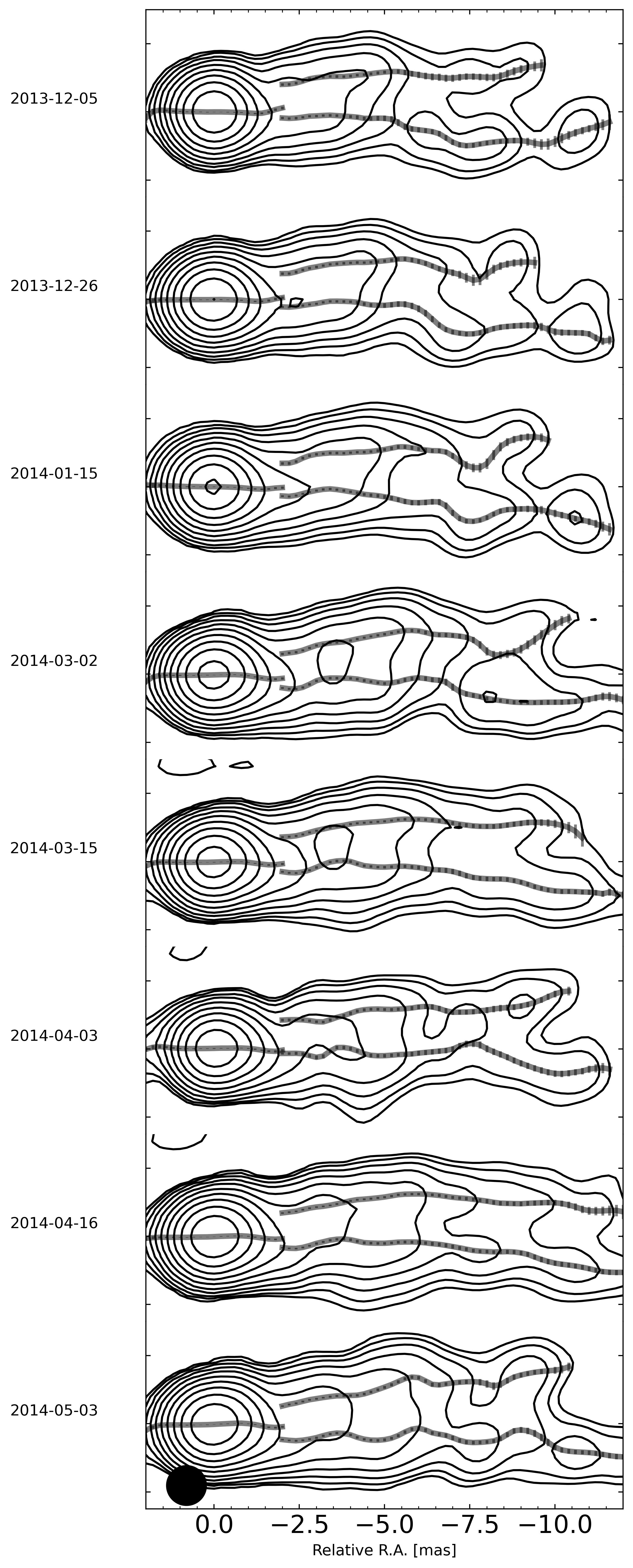}
    \includegraphics[width=0.3\textwidth]{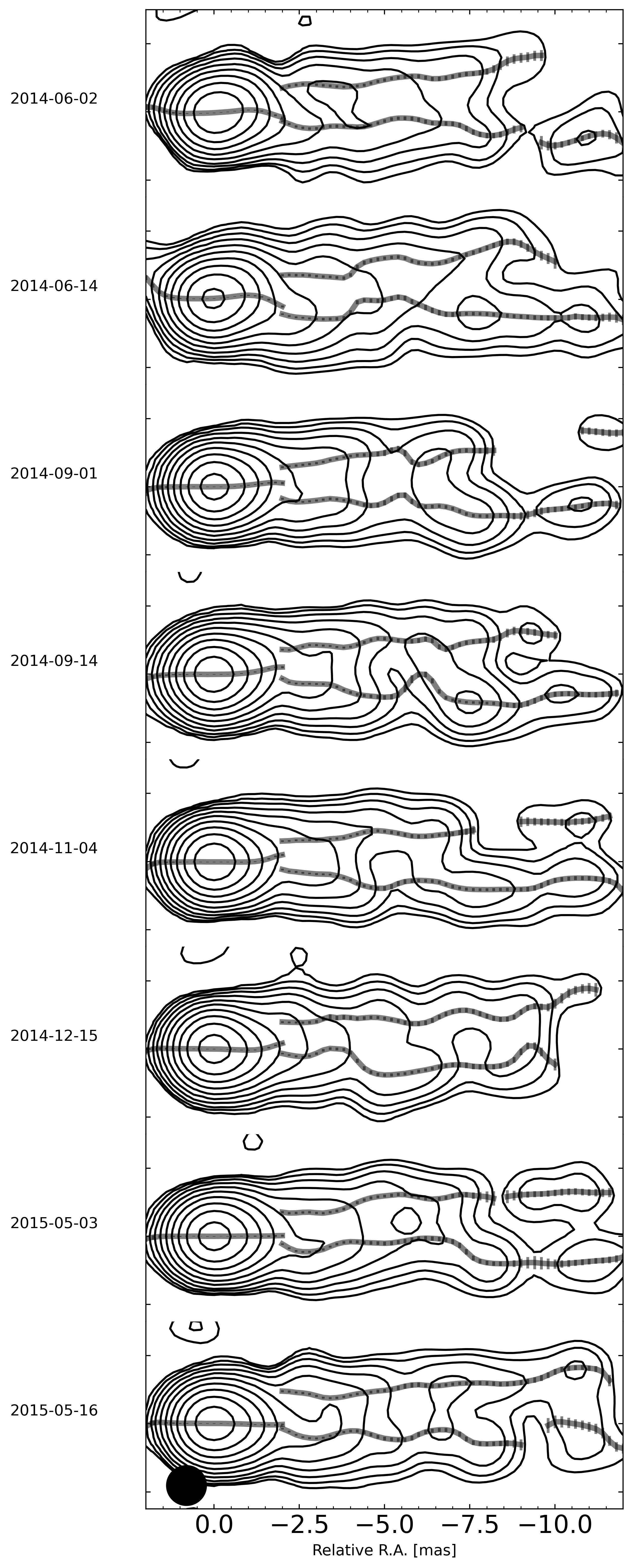}
    \includegraphics[width=0.3\textwidth]{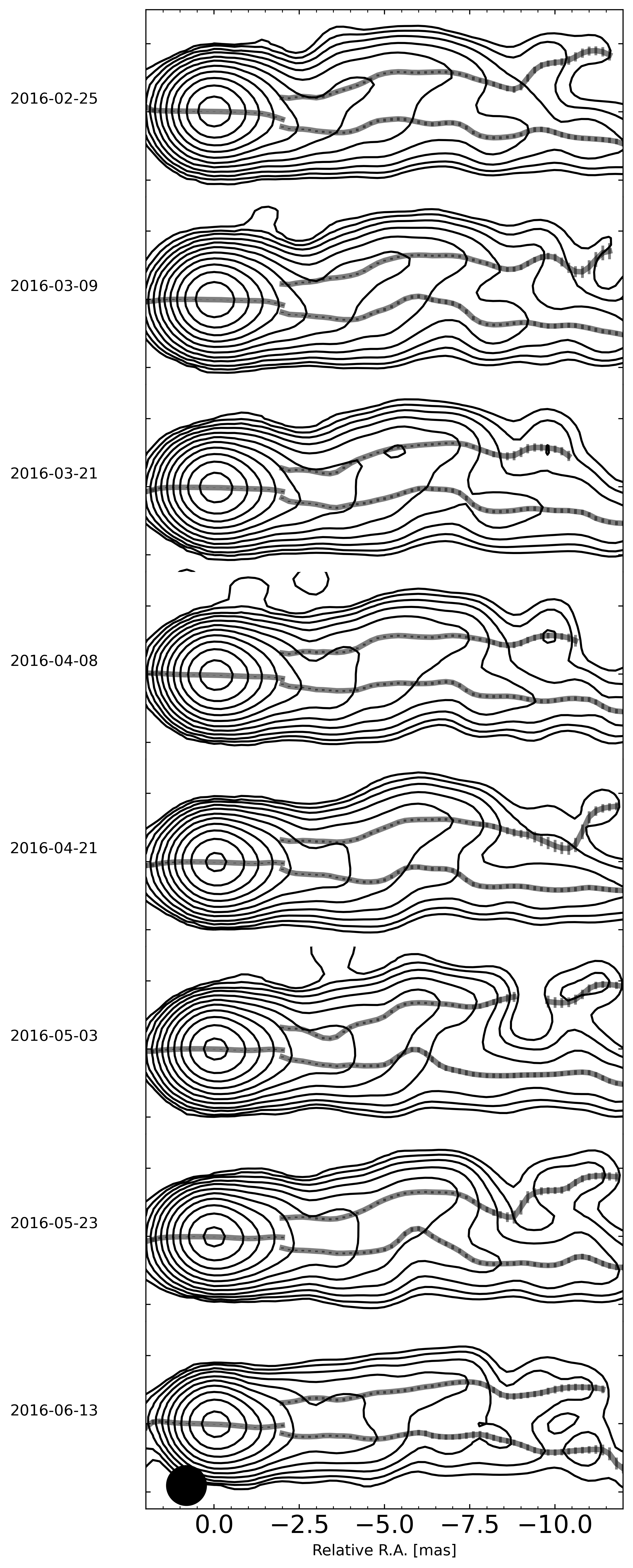}
    \caption{Ridge lines distribution of the M87 jet at 22\,GHz from KaVA observations in December 2013 to May 2014 (\textbf{left}), June 2014 to May 2015 (\textbf{middle}), and February 2016 to June 2016 (\textbf{right}). 
    The contours represent the total intensity. Contours start at 3$\sigma_{\text{rms}}$, increasing in steps of 2. The $\sigma_{\text{rms}}$ of each epoch is listed in Table~\ref{tab.a.1}. 
    The black circles in the bottom left corner represent the size of the restoring beam, namely 1.2\,mas circular Gaussian. All images have been rotated by $-$18\,degrees.}
    \label{fig:ridge_line_distributions}
\end{figure}

Figure~\ref{fig:ridge_line_distributions} 
shows the ridge lines of the M87 jet within 12\,mas from the core from the KaVA 22\,GHz observations between December 2013 to May 2014 (left panel), June 2014 to May 2015 (middle panel), and February 2016 to June 2016 (right panel), respectively.
Contours represent the total intensity, starting at 3$\sigma_{\text{rms}}$ and increasing by a factor of two.
Ridge lines that lie outside the contour lines or whose location changes abruptly are excluded.
The observation date of each datum is written on the left of the contours.
In general, the jet structures observed at KaVA 22\,GHz are well represented by double ridge lines up to 12\,mas from the core. 
The position error increases downstream. 
The northern limb tends to have shorter or more discontinuous ridge lines than the southern limb due to its weaker~intensity.

\section{Transverse Oscillation in the M87 Jet}\label{sec4}

\begin{figure}

    \includegraphics[width=0.49\textwidth]{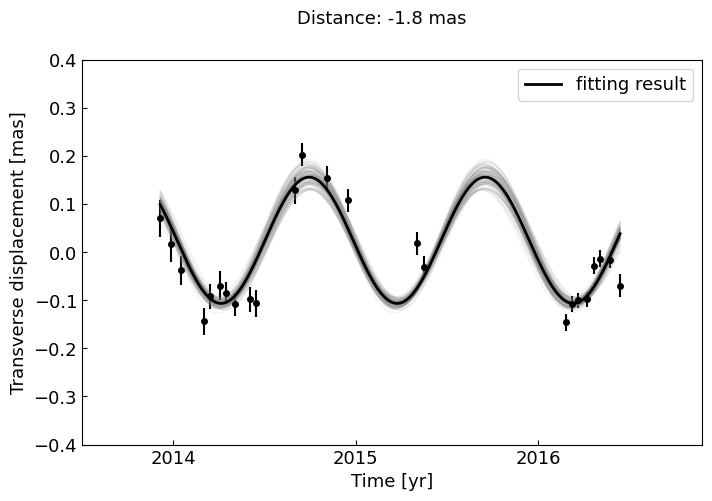}
    \includegraphics[width=0.49\textwidth]{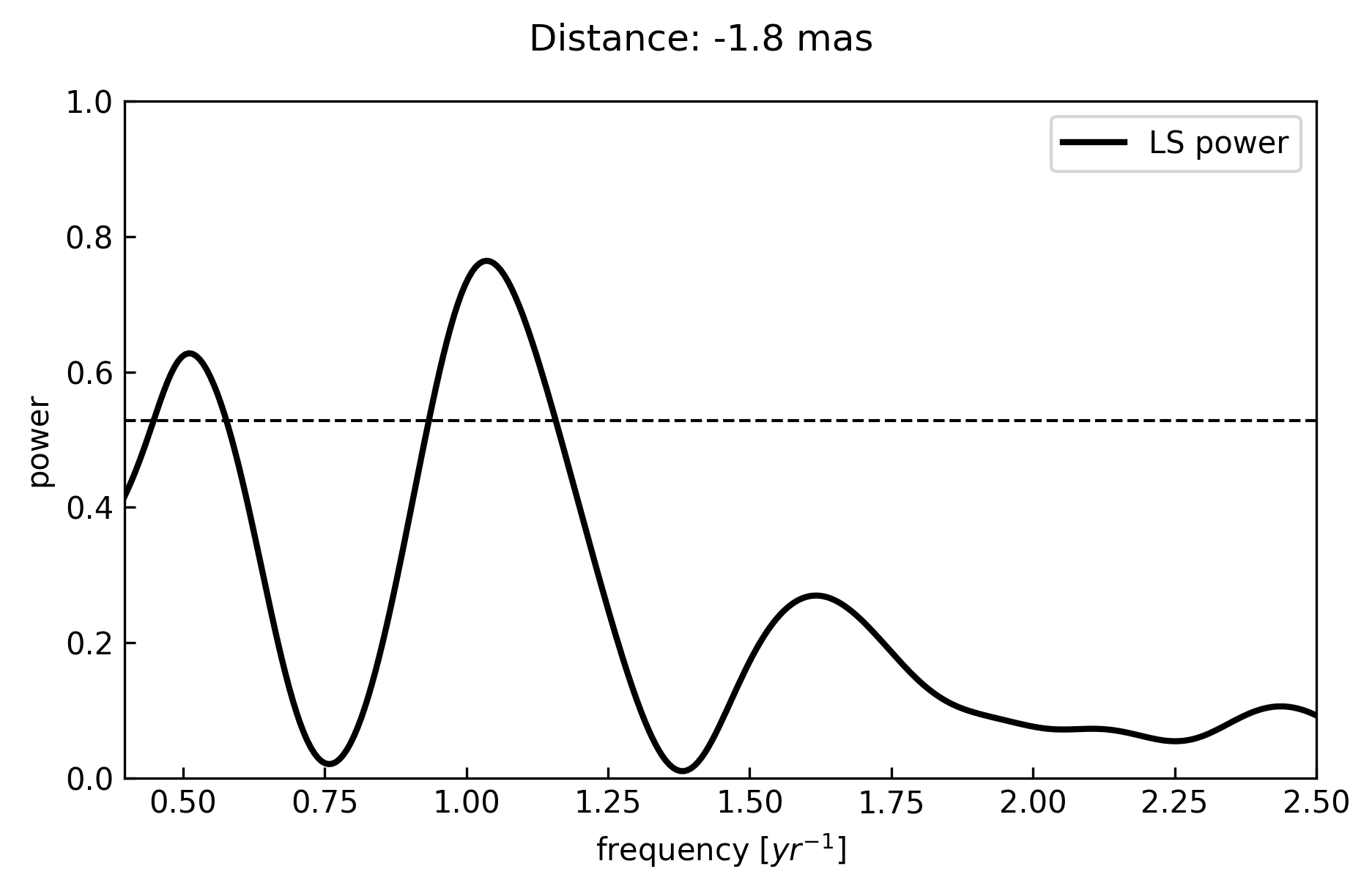}
    \includegraphics[width=0.49\textwidth]{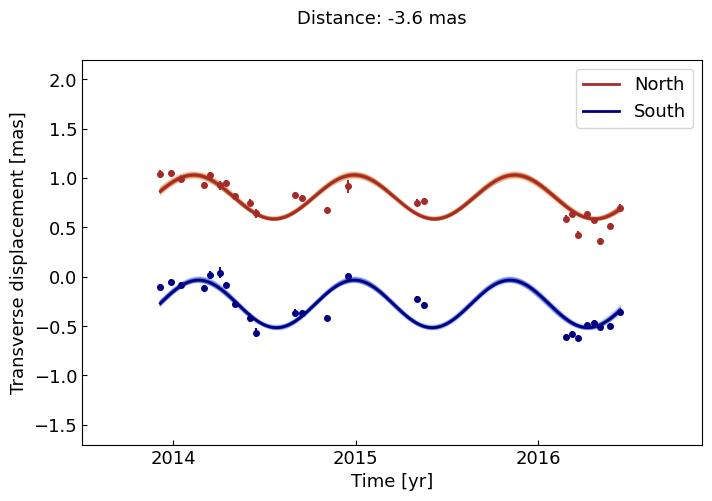}
    \includegraphics[width=0.49\textwidth]{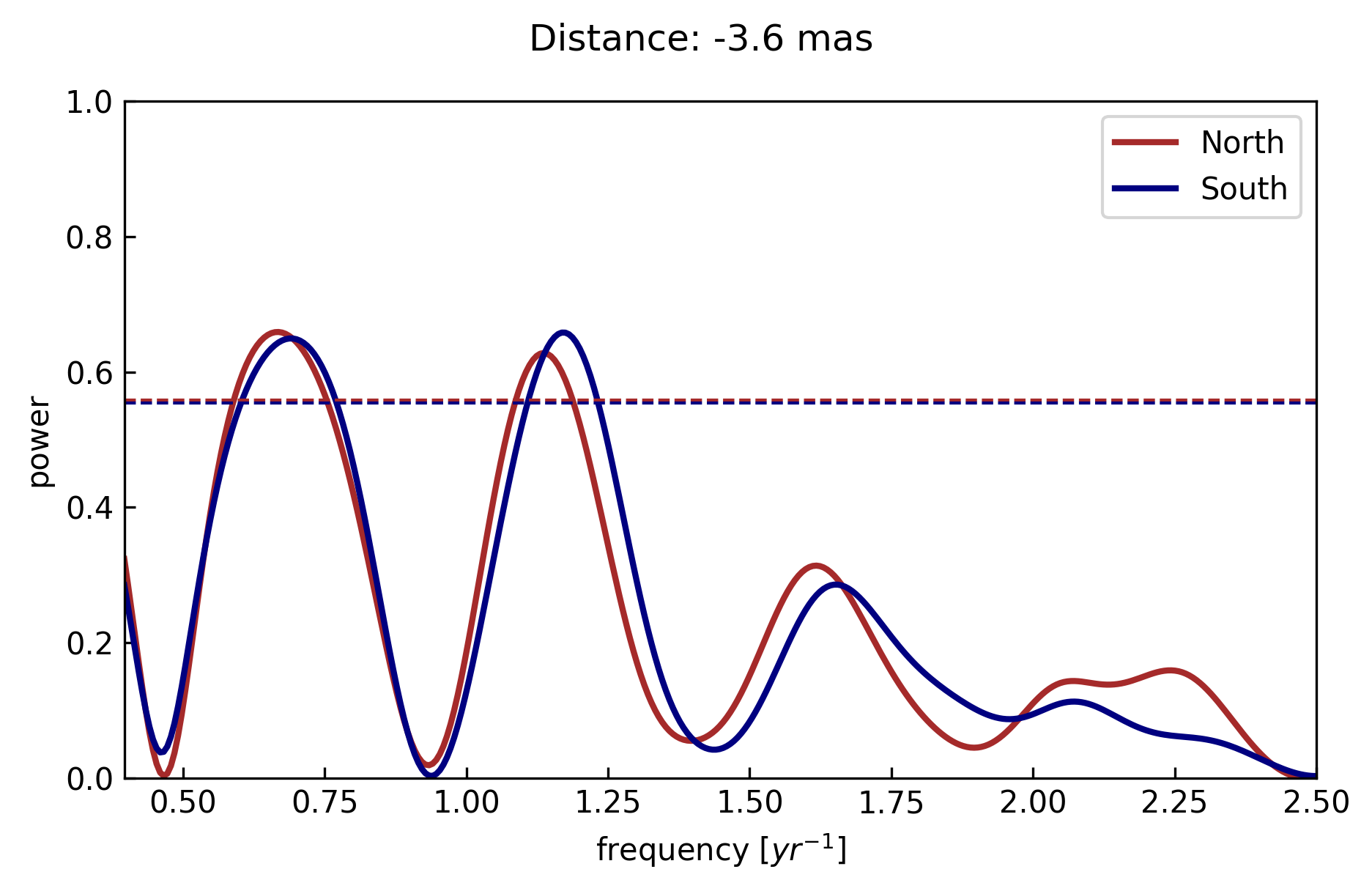}
    \includegraphics[width=0.49\textwidth]{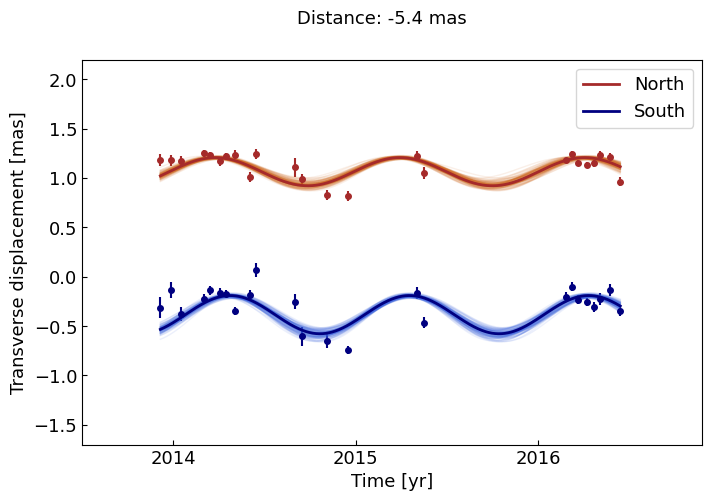}
    \includegraphics[width=0.49\textwidth]{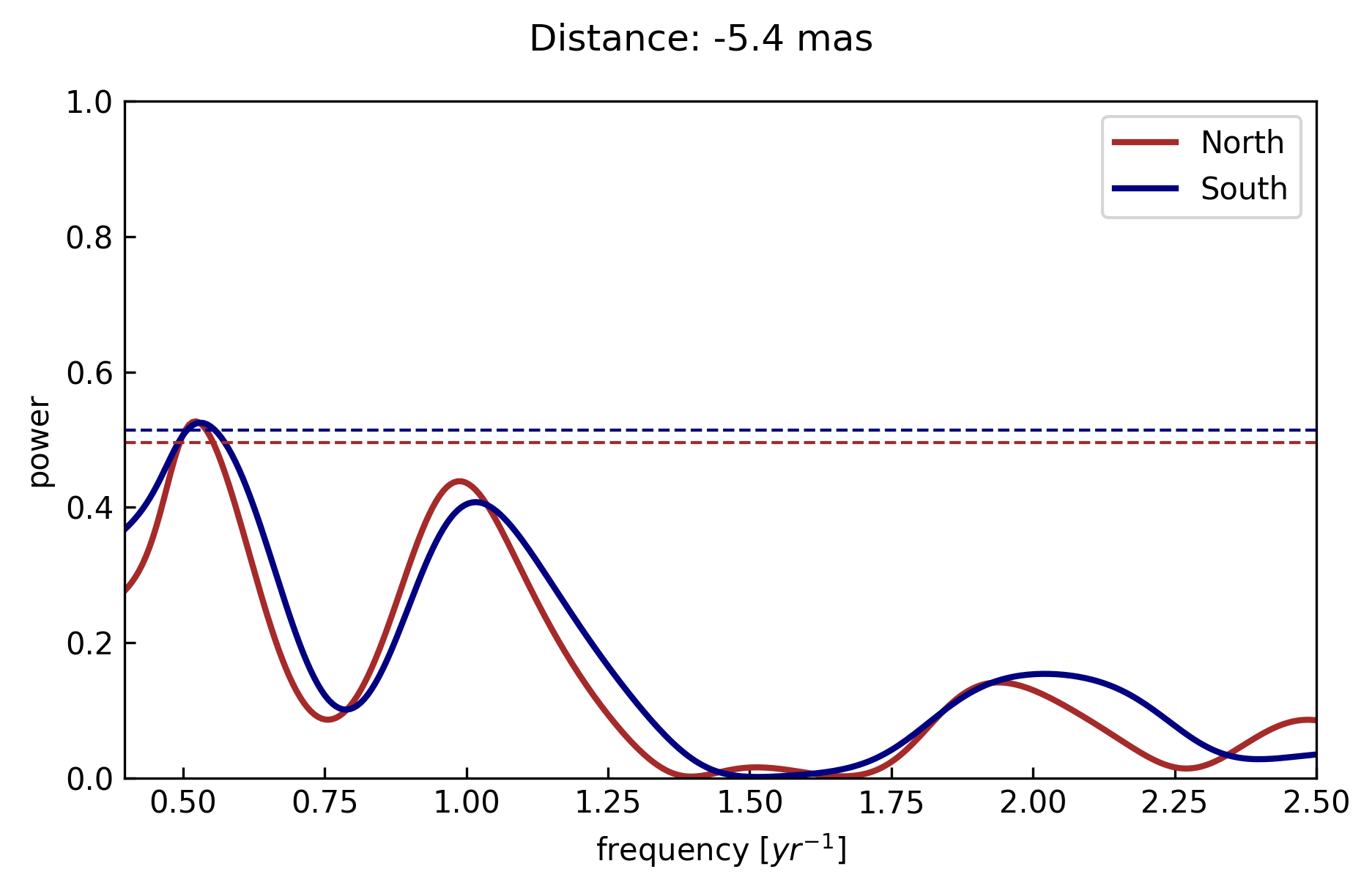}
    \includegraphics[width=0.49\textwidth]{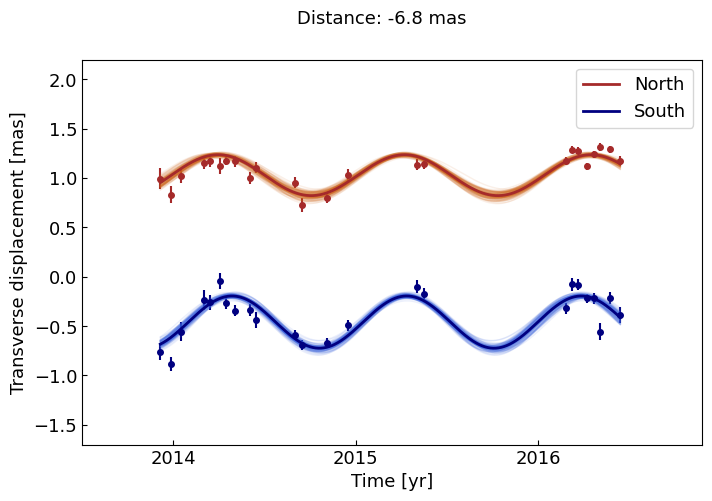}
    \includegraphics[width=0.49\textwidth]{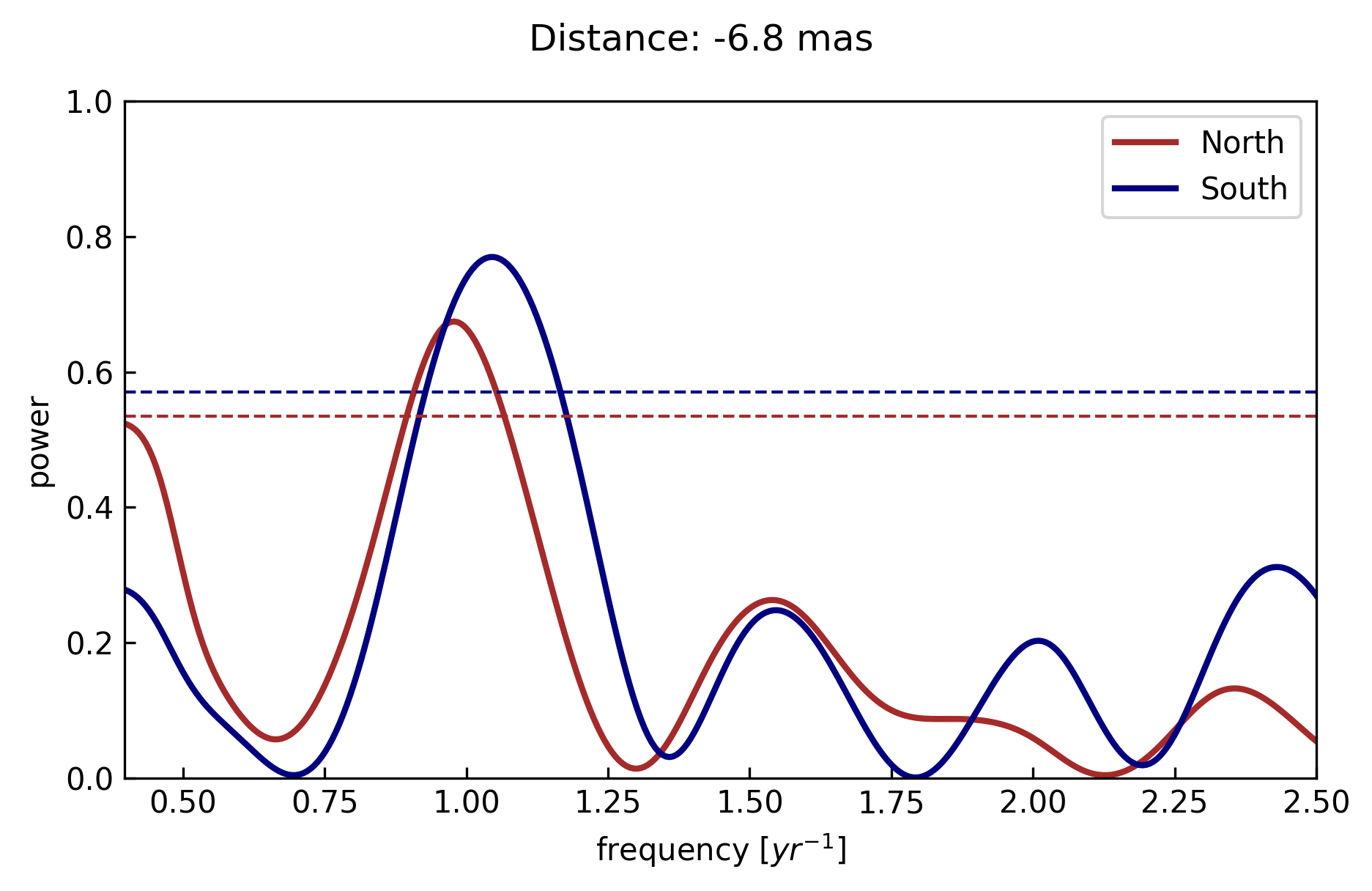}
    
    \caption{\textbf{Left:} Transverse oscillations of the M87 jet obtained at 1.8, 3.6, 5.4, and 6.8\,mas from the core from December 2013 to June 2016. The filled circle is the transverse displacement of the jet ridge from the jet axis. \replaced{The thick line is the best-fit sinusoidal model. The thin lines are 100 randomly chosen model parameters from the MCMC samples to show the statistical errors.}{The solid line shows that the least squares sinusoidal curve fits the data.} The black line is the result of the single ridge and the red and blue lines are the result of the northern and southern ridges, respectively. The amplitude ($A$), period ($P$), phase ($\phi$), and offset ($y_0$) of the sinusoidal fit are \replaced{summarized in Table~\ref{tab:sine_result}}{given in the figure}. \deleted{In all cases, the period is about one year.} \textbf{Right:} Lomb--Scargle periodogram results of the transverse oscillations of the M87 jet at the same locations. \added{The dashed horizontal line is the false alarm probability of 10\% with an assumption of white noise.}}\label{fig:sinusoidal_fitting}
\end{figure}

To examine the transverse evolution of the jet structure, we traced the changes of the ridge lines with time in the transverse direction.
The left panels of Figure~\ref{fig:sinusoidal_fitting} show the change in the transverse displacement of the ridge from the jet axis at distances of 1.8, 3.6, 5.4, and 6.8\,mas. 
The 1.8\,mas distance is in the single ridge region, and the rest are in the double ridge region. 
The transverse displacements of the \replaced{n}{N}orthern ridge and \replaced{s}{S}outhern ridge are expressed in red and blue dots, respectively.
At all locations, the transverse position of the ridge line moves up and down as if oscillating with time.

\begin{figure}[H]
    \includegraphics[width=0.6\textwidth]{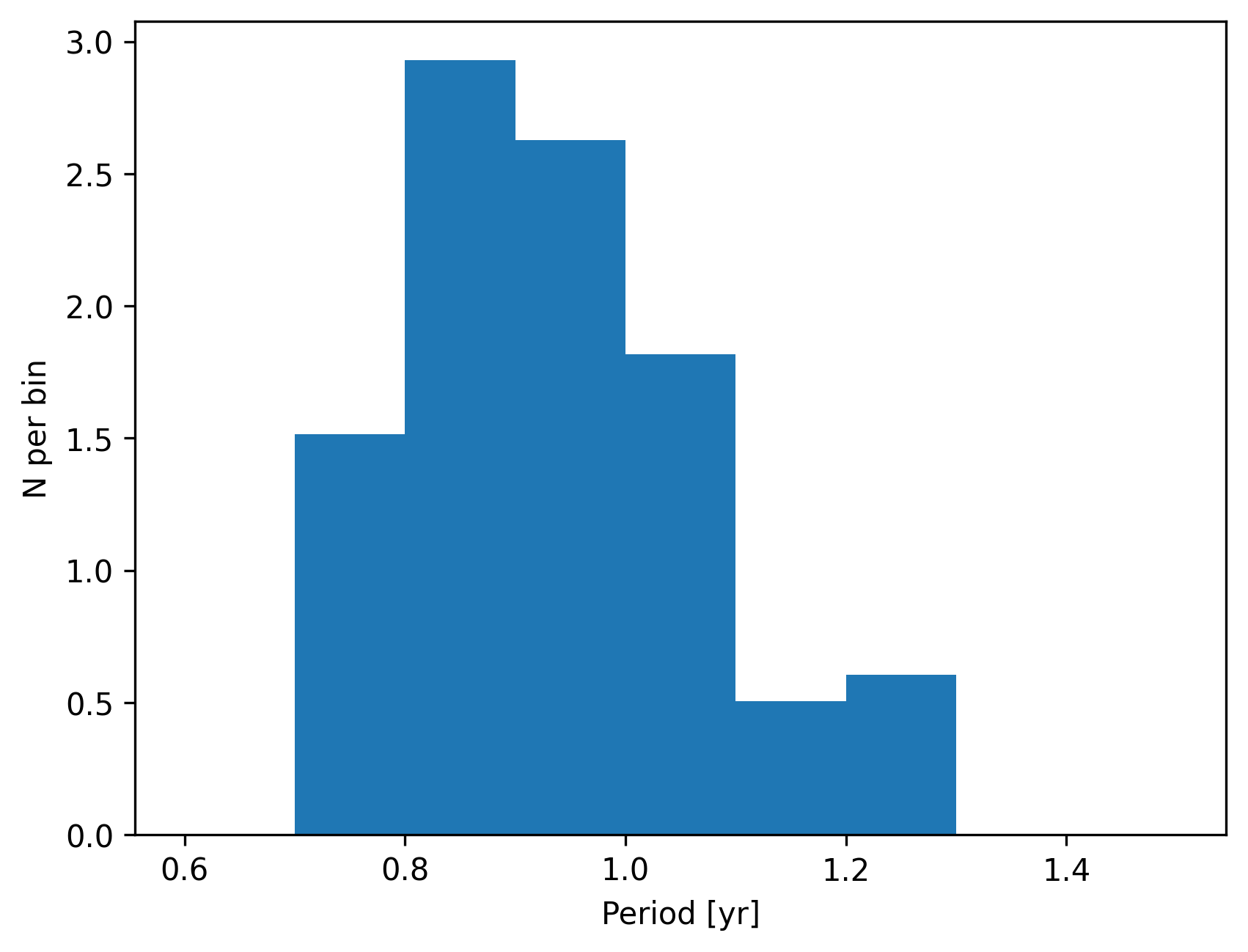}
    \caption{\added{Distribution of period of the transverse oscillations obtained from sinusoidal fit at all distances less than 12 mas. The average period is 0.94 yr, and the standard deviation is 0.12 yr.}}
    \label{fig:histogram_period}
\end{figure}

Several analyses were performed to investigate this transverse oscillatory motion. Firstly, we fitted the transverse displacement plot with a sinusoidal curve:
\begin{equation}
    y = A \sin\left(\replaced{\frac{2\pi}{P}t}{\frac{P}{2\pi}t} - \phi\right) + y_0
\end{equation} where $y$ is the transverse displacement, $t$ is time, $A$ is the amplitude, $P$ is the period, $\phi$ is the phase and $y_0$ is the offset of the sinusoidal curve.
\added{We performed Markov Chain Monte Carlo (MCMC) method to find the best-fit result, \added{e.g.,~\cite[][]{hogg2018}}.}
\deleted{The solid line is the least squares fit of the sinusoidal function to the data.}
\added{The thick line in the left panels of Figure~\ref{fig:sinusoidal_fitting} is the best sinusoidal fit result. The thin lines are 100 randomly chosen model parameters from the MCMC samples to show the statistical errors.}
The \replaced{best-fit parameters}{fitting results} are summarized in Table~\ref{tab:sine_result}.
We found that the transverse motion of the jet is well described by a sinusoidal curve, meaning that the jet oscillates periodically.
\deleted{Interestingly, the oscillation period is almost the same at all distances on both ridges and is around one year.}
\added{Figure~\ref{fig:histogram_period} is the distribution of periods obtained by sinusoidal fitting to the transverse motion of the jet at all distances less than 12 mas at both ridges. The average of the distribution is 0.94 yr, and the standard deviation is 0.12 yr.}
\added{This shows that the M87 oscillates transversely at all distances on both ridges with a similar period.}
We also found that the phase difference between the oscillations of the northern and southern limbs is less than one-tenth of the period. 
This means that the oscillations of the two ridges are almost in phase.
In Figure \ref{fig:beam_comparison}, we investigated whether the size of the restored beam could affect the sinusoidal fitting results, and found that the results are not changed.


\begin{table}[H] 
\caption{Summary of sinusoidal fit results for the transverse motion of the M87 jet \added{at a distance of 1.8, 3.6, 5.4, and 6.8}.\label{tab:sine_result}}
\newcolumntype{C}{>{\centering\arraybackslash}X}
\begin{tabularx}{\textwidth}{CCCCC}
\toprule
\added{\textbf{Distance}} & \textbf{1.8 mas} & \textbf{3.6 mas}	& \textbf{5.4 mas}	& \textbf{6.8 mas}\\
\midrule
$A$ (mas)    & \replaced{0.131~$\pm$~0.008}{0.132~$\pm$~0.008}   & 0.223~$\pm$~0.010                                & \replaced{0.143~$\pm$~0.017}{0.145~$\pm$~0.017}	    & \replaced{0.207~$\pm$~0.019}{0.209~$\pm$~0.019} \\
$P$ (yr)     & \replaced{0.966~$\pm$~0.010}{0.97~$\pm$~0.01}     & \replaced{0.880~$\pm$~0.005}{0.88~$\pm$~0.01}      & \replaced{1.013~$\pm$~0.020}{0.88~$\pm$~0.01}	    & \replaced{1.023~$\pm$~0.017}{1.02~$\pm$~0.02} \\
$\phi$ (rad) & \replaced{3.742~$\pm$~0.105}{3.74~$\pm$~0.11}     & \replaced{6.019~$\pm$~0.064}{6.02~$\pm$~0.06}      & \replaced{0.302~$\pm$~0.166}{0.30~$\pm$~0.17}	    & \replaced{0.371~$\pm$~0.198}{0.37~$\pm$~0.15} \\
$y_0$ (mas)  & 0.025~$\pm$~0.007                               & \replaced{0.808~$\pm$~0.007}{0.800~$\pm$~0.007}    & \replaced{1.064~$\pm$~0.015}{1.06~$\pm$~0.01}	    & \replaced{1.029~$\pm$~0.015}{1.02~$\pm$~0.015} \\
$A$ (mas)    &       -                                       & 0.241~$\pm$~0.010	                              & \replaced{0.194~$\pm$~0.020}{0.196~$\pm$~0.020}	    & \replaced{0.264~$\pm$~0.017}{0.265~$\pm$~0.017} \\
$P$ (yr)     &       -                                       & \replaced{0.854~$\pm$~0.004}{0.85~$\pm$~0.00}	  & \replaced{0.983~$\pm$~0.018}{0.98~$\pm$~0.02}	    & \replaced{0.958~$\pm$~0.014}{0.96~$\pm$~0.01} \\
$\phi$ (rad) &       -                                       & \replaced{6.256~$\pm$~0.064}{6.26~$\pm$~0.06}	  & \replaced{0.878~$\pm$~0.168}{0.87~$\pm$~0.17}       & \replaced{1.006~$\pm$~0.147}{1.02~$\pm$~0.14} \\
$y_0$ (mas)  &       -                                       & $-$0.275~$\pm$~0.007                               & \replaced{$-$0.384~$\pm$~0.017}{-0.387~$\pm$~0.017}   & $-$0.459~$\pm$~0.015\\
\bottomrule
\end{tabularx}
\end{table}

We also performed Lomb--Scargle periodogram analysis for each transverse oscillation plot~\cite{lomb76, scargle82}.
We evaluate the periodogram power at frequencies between $f_{\text{min}}=1/T$, where $T$ is the total time of the observations and $f_{\text{max}}=1/(2T_{\text{min}})$, corresponding to one over twice the minimum time interval between observations~\cite{vanderplas18}. 
The number of grids in the frequency range was set to 2,000 in order to sufficiently sample at every period.
The results of the periodogram analyses at each distance are shown in the right panels of Figure~\ref{fig:sinusoidal_fitting}.
\replaced{We found that the periodogram shows a strong peak at a similar place as the value obtained from the sinusoidal fit.}{We found that the periodogram shows a strong peak at $\sim1$ year for all distances. This peak is consistent with the sinusoidal fit results.}
We also found another peak at a lower frequency, and the power of this \added{low-frequency} peak is similar or slightly stronger than the \replaced{high-frequency}{one-year} peak at 3.6 and 5.4\,mas.
However, the position of the \replaced{low-frequency}{second} peak changes with distance and even disappears at 6.8\,mas.
Therefore, it is not caused by the oscillations inside the jet and may be an artifact caused by the non-uniform time interval of the observations, e.g.,~\cite[][]{vanderplas18}.
Further observations may clarify the presence of the low-frequency peak.
\added{The horizontal line stands for the false-alarm probability (i.e., the probability that a signal with no periodic component would lead to a peak of this magnitude) of 10\% with an assumption of white noise} \endnote{That is, the power of the  random component in the periodogram is constant with respect to the frequency. In many situations of interest, however, this assumption does not hold. For more discussions, see \citet{vanderplas18} and references therein.} \cite[][]{vanderplas18}. \added{This shows that the probability of which the periodic signal originated from the random component is less than 10\% except at 5.4 mas.}


From these analyses, we were able to reveal transverse oscillations of the M87 jet \added{at all distances} with a\added{n} \added{average} period of \replaced{$0.94\pm0.12$}{$\sim1$} year \deleted{at all distances}.
This fast and periodic oscillation in the M87 jet has never been reported in previous studies.
The period is much shorter than the quasi-periodic sideways shifts reported by \citet{walker2018} (8--10 years).
It is relatively similar to the timescale of the transition of the jet structures reported by \citet{britzen2017} ($\sim$~2\,years), but still about half of it.
Therefore, it is possible that either we discovered a totally new type of oscillation with a much faster period or more clearly revealed oscillations that previous studies could not fully track due to their insufficient time sampling.

\subsection*{Origins of the Transverse Oscillations}
Given that the period of the oscillations is about one year, it is a natural question whether the oscillation originates from the projection effect of the Earth's revolution. 
However, the parallax angle of the M87 is $\sim$~0.1\,$\mu$as, which is 1000 times smaller than the observed oscillation amplitude ($\sim$~0.1 mas). 
Thus, we can rule out the possibility that the observed oscillations originate from Earth's parallax.
Future observations are able to test this possibility, as the Earth's revolution predicts perfectly periodic oscillations with an exactly one-year period.

Several jet-intrinsic origins can be proposed to explain transverse oscillations in the M87 jet.
\replaced{Advection}{Propagation} of the distorted jet structure created by magneto-hydrodynamic (MHD) instabilities is one of the possible scenarios, e.g.,~\cite[][]{hardee11}. 
Several previous studies have suggested Kelvin--Helmholtz instability (KHI) to describe the M87 \added{jet}'s twisted filament structures, which are shown up to kpc-scales, e.g.,~\cite[][]{lobanov03, hardee-eilek2011, pasetto21}. 
\citet{walker2018} explained the long-term sideways shifts observed in pc-scale jets as the propagation of a helical pattern created by KHI along the jet.
However, it is unclear whether the newly discovered fast oscillation can be described by KHI, as the region observed by KaVA may be highly magnetized, e.g.,~\cite[][]{kino2015}. 
Rather, current-driven instability (CDI) such as sausage instability (m = 0) or kink instability (m = 1) can cause oscillations in the magnetically dominated regions e.g.,~\cite[][]{mizuno09, singh16, dong20}.

\added{For a highly magnetized jet, the propagation of transverse MHD waves (i.e., Alfv\'{e}n waves) is another possible scenario to explain the transverse oscillation~\cite{cohen15}. This scenario requires a source to excite the wave upstream of the jet, in analogy to exciting a wave on a whip by shaking the handle, and the jet's magnetic field strength and mass density determine the speed of wave propagation.}

{Alternatively, MHD simulations of the magnetically arrested disk (MAD) have shown that oscillations in the energy outflow of the jet could be induced by the accumulation and escape of magnetic field line bundles in the innermost part of accretion disks~\cite{tchekhovskoy2011, mckinney2012}. 
According to this scenario, multiple harmonics of the quasi-periodic oscillations in the jet are expected~\cite{mckinney2012}. 
Future observations can test this scenario by looking for the additional oscillations in the jet.
}

\section{Summary}\label{sec5}

In this work, we found a transverse oscillation of the M87 jet with a\added{n average} period of \replaced{$0.94\pm0.12$ years}{$\sim1$ year} using high-cadence monitoring observations of KaVA at 22\,GHz. 
It has a much shorter period than that reported in previous studies.
Therefore, it may be a new type of oscillation that has never been reported or is not clearly seen in previous observations due to insufficient time intervals.
The oscillation has almost the same period at all positions of the jet and has an amplitude of about $\sim$~0.1\,mas. 
The amplitude of the oscillation is about 1000 times larger than the amplitude expected from Earth's parallax ($\sim$~0.1\,$\mu$as), which excludes the possibility that this oscillation originated from the Earth revolution.
We found that the oscillations were present in both the north and south limbs in nearly the same phase.
We proposed that the propagation of jet instability or perturbed mass accretion occurred in the magnetically arrested disks as possible origins.

Since 2017, KaVA has expanded as EAVN (East Asia VLBI Network;~\cite{cui21}) by joining Chinese telescopes. 
EAVN monitoring is currently the only monitoring that steadily observes the M87 jet.
Using these observations, we will investigate the transverse oscillations in the M87 jet in more detail.

\vspace{6pt} 



\authorcontributions{{
Conceptualization, H.R., M.K. and K.H.; 
methodology, H.R., M.K. and K.H.; 
software, H.R.; 
validation, H.R., K.Y. and Y.C.; 
formal analysis, H.R.; 
data curation, H.R., K.Y., Y.C. and K.H.; 
writing---original draft, H.R.; 
writing---review and editing, K.Y., Y.C., M.K., K.H., T.K., Y.M., B.W.S. and F.T.; 
visualization, H.R.; 
supervision, M.K. and B.W.S.; 
project administration, M.K. and B.W.S.; 
funding acquisition, B.W.S.
All authors have read and agreed to the published version of the manuscript.}
}

\funding{This work is funded by the following:
This research was supported by the Korea Astronomy and Space Science Institute under the R\&D program supervised by the Ministry of Science and ICT.
H.R. and B.W.S. acknowledge support from the  KASI-Yonsei DRC program of the Korea Research Council of Fundamental Science and Technology (DRC-12-2-KASI).
Y.C. is supported by the China Postdoctoral Science Foundation ({2022M712084}). 
And just in case, please also make sure my grant no. is 2022M712084.
Y.M. is supported by the National Natural Science Foundation of China (12273022) and the Shanghai pilot program of international scientists for basic research (22JC1410600).
} 

\dataavailability{{All observing data obtained by KaVA, except the data in the term of the right of occupation by a principal investigator, are archived via the following website. 
\\
\url{https://radio.kasi.re.kr/arch/search.php}
}
}

\acknowledgments{This work is based on observations made with the KaVA, which is operated by the Korea Astronomy and Space Science Institute and the National Astronomical Observatory of Japan.}

\conflictsofinterest{{The authors declare no conflicts of interest.}
} 





\appendixtitles{yes} 
\appendixstart
\appendix
\section[\appendixname~\thesection]{Image Information of KaVA 22\,GHz Monitoring of the M87 Jet from December 2013 to June 2016}
\begin{figure}[H]
    \includegraphics[width=0.9\textwidth]{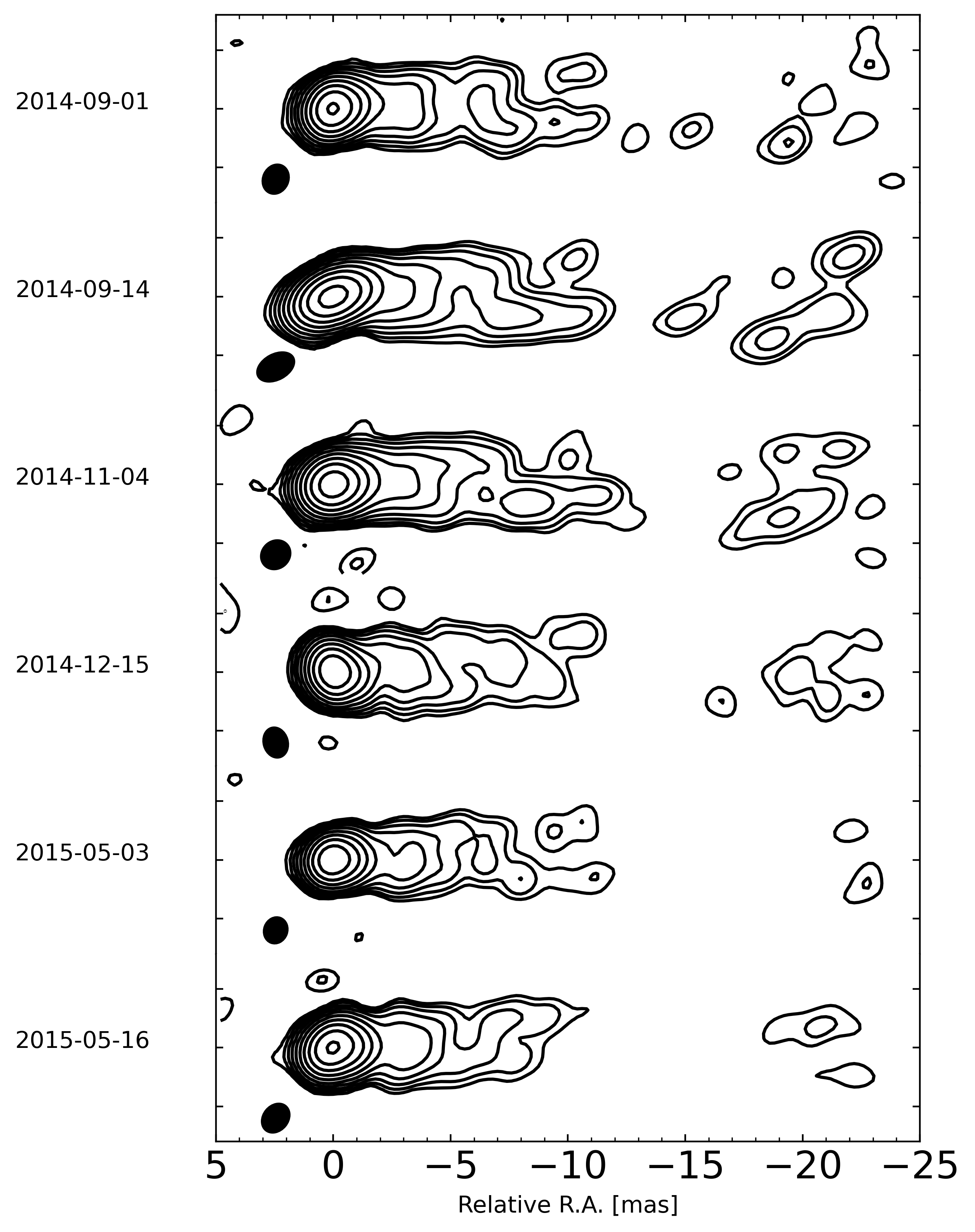}
    \caption{Total intensity maps of KaVA 22\,GHz observations of the M87 jet between September 01 2014 and May 16 2015. Contours start at 3$\sigma_{\text{rms}}$, increasing in steps of 2. The synthesized beam for individual epochs is drawn as a black ellipse on the bottom left side of the individual maps.}
    \label{fig:total_intensity_map}
\end{figure}

\begin{table}[H] 
\tablesize{\footnotesize }
\caption{Summary of KaVA 22\,GHz observations of the M87 jet from December 2013 to June 2016.\label{tab.a.1}}
\newcolumntype{C}{>{\centering\arraybackslash}X}
\begin{tabularx}{\textwidth}{C C C C C}
\toprule
\textbf{Obs. Date}	& \textbf{Beam Size} & \textbf{$I_{\text{peak}}$} & \textbf{$\sigma_{\text{rms}}$} & \textbf{Dynamic Range}\\
 &\textbf{ (mas$\times$mas,~deg)} &\textbf{ (Jy/beam)} & \textbf{(mJy/beam) }& \\
\textbf{ (1) }&\textbf{ (2) }&\textbf{ (3) }& \textbf{(4) }&\textbf{ (5)} \\
\midrule
{5 December 2013} 
  &   1.31~$\times$~1.14, $-$18.0   &   1.67 &   1.003     &  1669 \\
{26 December 2013}  &   1.61~$\times$~1.23, $-$42.7   &   1.32 &   0.784     &  1688 \\
{15 January 2014} &   1.83~$\times$~1.51, $-$54.3   &   1.36 &   0.76     &  1794 \\
{2 March 2014}  &   1.50~$\times$~1.21, 3.0   &   1.39 &   0.516     &  2695 \\
{15 March 2014}  &   1.34~$\times$~1.14, $-$15.9   &   1.42 &   0.464     &  3055 \\
{3 April 2014}  &   1.32~$\times$~1.22, 14.5   &   1.35 &   0.775     &  1742 \\
{16 April 2014}  &   1.27~$\times$~1.10, $-$17.7   &   1.39 &   0.37     &  3758 \\
{3 May 2014}  &   1.47~$\times$~1.26, 11.8   &   1.42 &   0.386     &  3672 \\
{2 June 2014}  &   1.08~$\times$~0.91, $-$18.3  &   1.17 &   0.734     &  1600 \\
{14 June 2014}  &   1.34~$\times$~1.10, $-$11.5   &   1.11 &   0.531     &  2085 \\
{1 September 2014}  &   1.35~$\times$~1.16, $-$7.9   &   1.27 &   0.627     &  2024 \\
{14 September 2014}  &   1.56~$\times$~1.15, $-$41.3  &   1.28 &   0.497     &  2580 \\
{4 November 2014} &   1.41~$\times$~1.26, $-$32.2  &   1.32 &  0.491     &  2695 \\
{15 December 2014}  &   1.27~$\times$~1.10, 2.34  &   0.92 &   0.442     &  2077 \\
{3 May 2015}  &   1.22~$\times$~1.08, $-$4.1  &   1.25 &   0.603     &  2070 \\
{16 May 2015} &   1.42~$\times$~1.15, $-$21.5  &   1.26 &   0.515     &  2448 \\
{25 February 2016}  &   1.31~$\times$~1.15, $-$5.9  &   1.42 &   0.289     &  4903 \\
{9 March 2016}  &   1.38~$\times$~1.15, $-$2.3  &   1.35 &   0.247     &  5474 \\
{21 March 2016}  &   1.55~$\times$~1.26, $-$14.9  &   1.40 &   0.291     &  4825 \\
{8 April 2016}  &   1.36~$\times$~1.12, 1.9  &   1.27 &   0.287     &  4439 \\
{21 April 2016}  &   1.28~$\times$~1.16, $-$3.7 &   1.29 &   0.357     &  3605 \\
{3 May 2016} &   1.30~$\times$~1.05, $-$11.9  &   1.07 &   0.281     &  3794 \\
{23 May 2016}  &   1.24~$\times$~1.08, $-$12.6  &   1.12 &   0.303     &  3690 \\
{13 June 2016}  &   1.26~$\times$~1.14, 5.3  &   0.79 &   0.394     &  2013 \\

\bottomrule
\end{tabularx}
NOTE---(1) Observation date. (2) Synthesized beam size with a natural weighting scheme. (3) Peak intensity of the image with a natural weighting scheme. (4) Image rms noise ($\sigma_{\text{rms}}$) with a natural weighting scheme. (5) Dynamic range of the image calculated from $I_{\text{peak}}$/$\sigma_{\text{rms}}$.
\end{table}

\section[\appendixname~\thesection]{\added{Effect of the Restoring Beam Size}}
\begin{figure}[H]
  
    \includegraphics[width=0.44\textwidth]{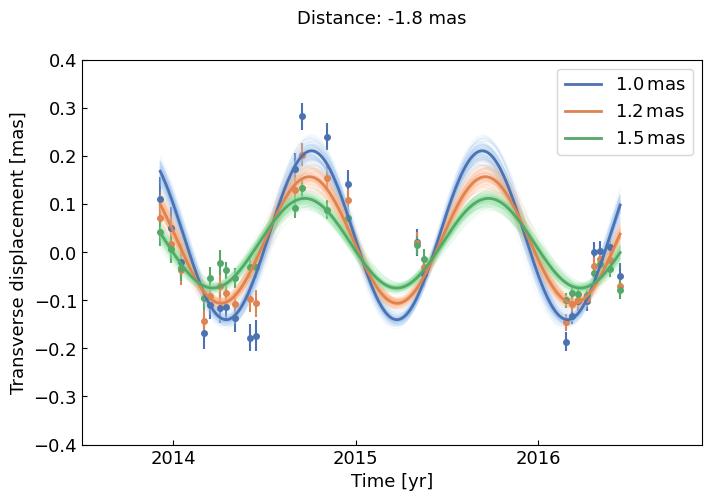}
    \includegraphics[width=0.44\textwidth]{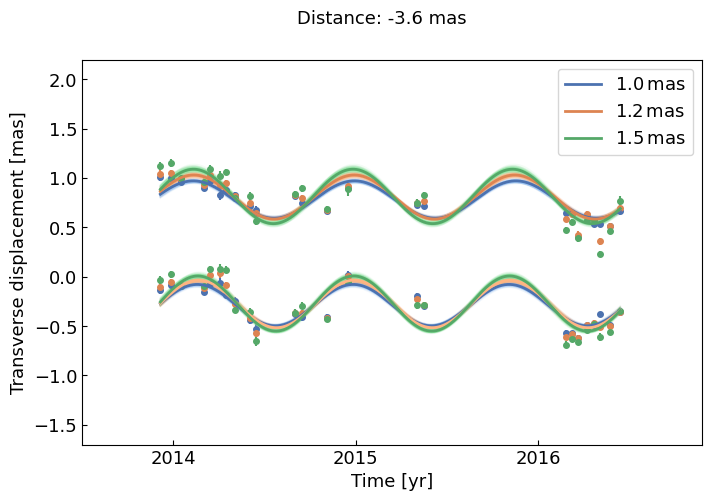}\\
    \includegraphics[width=0.44\textwidth]{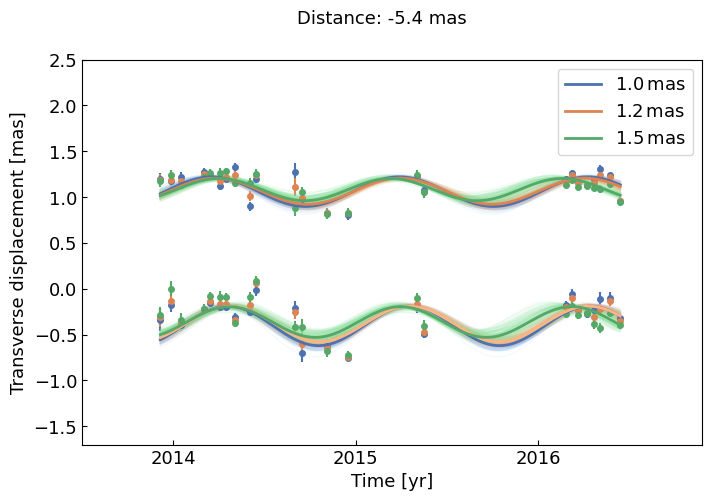}
    \includegraphics[width=0.44\textwidth]{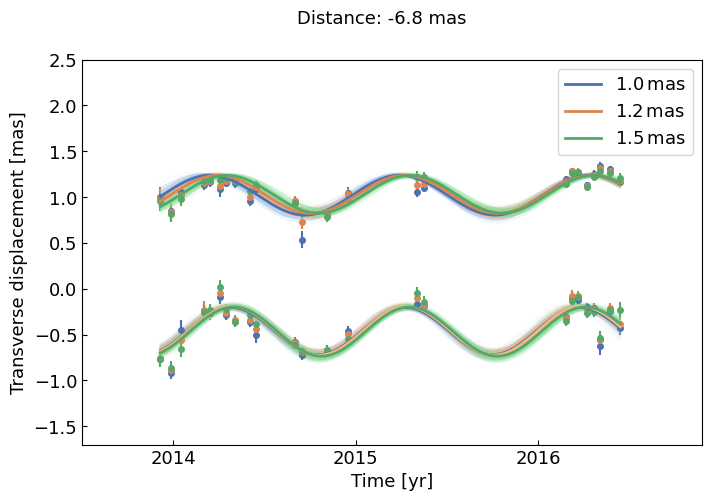}
    \caption{\added{Transverse oscillations of the M87 jet, obtained at distances of 1.8, 3.6, 5.4, and 6.8 mas by applying different sizes of restoring beams. The blue, orange, and green dots represent the transverse displacement of the ridge lines obtained by applying circular restoring beams with a radius of 1.0, 1.2, and 1.5 mas, respectively, and the solid lines are their sinusoidal fitting results. At all distances, the oscillation period is almost unchanged with respect to the beam size, while the amplitude is slightly~different.}}
    \label{fig:beam_comparison}
\end{figure}

\added{\textls[-15]{In this work, we used the restoring beam of 1.2~$\times$~1.2 mas while making ridge lines. However, since the separation between the northern limb and the southern limb is not significant, the choice of the restoring beam size could affect the oscillation analysis. 
Figure~\ref{fig:beam_comparison} }summarizes the transverse motion of the ridge lines and the sinusoidal fitting results obtained using different restoring beams at distances of 1.8, 3.6, 5.4 and 6.8 mas.
Different colors represent different beam sizes, including smaller (1.0~$\times$~1.0~mas) and larger (1.5~$\times$~1.5~mas) beams.
From these experiments, we found that the size of the restoring beam does not affect the oscillation period, while there are some differences in amplitude, especially in the upstream region where the gap between the northern and southern ridge is smaller. It confirms that the main conclusion of this study is not affected by the beam size.}


\begin{adjustwidth}{-\extralength}{0cm}
\printendnotes[custom] 

\reftitle{References}

\PublishersNote{}

\end{adjustwidth}
\end{document}